\documentstyle[referee]{mn}
\newif\ifAMStwofonts

\input epsf
  \ifCUPmtlplainloaded \else
  \fi
  \ifAMStwofonts
    \ifCUPmtlplainloaded \else
      \NewSymbolFont{upmath} {eurm10}
      \NewSymbolFont{AMSa} {msam10}
      \NewMathSymbol{\upi}     {0}{upmath}{19}
      \NewMathSymbol{\umu}     {0}{upmath}{16}
      \NewMathSymbol{\upartial}{0}{upmath}{40}
      \NewMathSymbol{\leqslant}{3}{AMSa}{36}
      \NewMathSymbol{\geqslant}{3}{AMSa}{3E}

      \let\leq=\leqslant 
      \let\geq=\geqslant 
    \fi
  \fi
\ifnfssone
  \newmathalphabet{\mathit}
  \addtoversion{normal}{\mathit}{cmr}{m}{it}
  \addtoversion{bold}{\mathit}{cmr}{bx}{it}
  \newmathalphabet{\mathbfit} 
  \addtoversion{normal}{\mathbfit}{cmr}{bx}{it}
  \addtoversion{bold}{\mathbfit}{cmr}{bx}{it}
  \newmathalphabet{\mathbfss} 
  \addtoversion{normal}{\mathbfss}{cmss}{bx}{n}
  \addtoversion{bold}{\mathbfss}{cmss}{bx}{n}
  \ifAMStwofonts
    \ifCUPmtlplainloaded \else
      \UseAMStwoboldmath
      \makeatletter
      \new@mathgroup\upmath@group
      \define@mathgroup\mv@normal\upmath@group{eur}{m}{n}
      \define@mathgroup\mv@bold\upmath@group{eur}{b}{n}
      \edef\UPM{\hexnumber\upmath@group}
      \new@mathgroup\amsa@group
      \define@mathgroup\mv@normal\amsa@group{msa}{m}{n}
      \define@mathgroup\mv@bold\amsa@group{msa}{m}{n}
      \edef\AMSa{\hexnumber\amsa@group}
      \makeatother
      \mathchardef\upi="0\UPM19
      \mathchardef\umu="0\UPM16
      \mathchardef\upartial="0\UPM40
      \mathchardef\leqslant="3\AMSa36
      \mathchardef\geqslant="3\AMSa3E

      \let\leq=\leqslant 
      \let\geq=\geqslant 
    \fi
  \fi
\fi 

\ifnfsstwo
  \DeclareMathAlphabet{\mathbfit}{OT1}{cmr}{bx}{it}
  \SetMathAlphabet\mathbfit{bold}{OT1}{cmr}{bx}{it}
  \DeclareMathAlphabet{\mathbfss}{OT1}{cmss}{bx}{n}
  \SetMathAlphabet\mathbfss{bold}{OT1}{cmss}{bx}{n}
  \ifAMStwofonts
    \ifCUPmtlplainloaded \else
      \DeclareSymbolFont{UPM}{U}{eur}{m}{n}
      \SetSymbolFont{UPM}{bold}{U}{eur}{b}{n}
      \DeclareSymbolFont{AMSa}{U}{msa}{m}{n}
      \DeclareMathSymbol{\upi}{0}{UPM}{"19}
      \DeclareMathSymbol{\umu}{0}{UPM}{"16}         			
      \DeclareMathSymbol{\upartial}{0}{UPM}{"40}
      \DeclareMathSymbol{\leqslant}{3}{AMSa}{"36}
      \DeclareMathSymbol{\geqslant}{3}{AMSa}{"3E}

      \let\leq=\leqslant 
      \let\geq=\geqslant 
    \fi
  \fi
\fi 

\ifCUPmtlplainloaded \else
  \ifAMStwofonts \else 
    \def\upi{\pi}
    \def\umu{\mu}
    \def\upartial{\partial}
  \fi
\fi

\title{The role of gravitational supernovae on the galactic evolution of 
the LiBeB isotopes}

\author[Carlos Abia, Jordi Isern and Ute Lisenfeld] 
{Carlos Abia$^1$, Jordi Isern$^2$, Ute Lisenfeld$^3$\\
  $^1$ Departamento de F\'\i sica Te\'orica y del Cosmos, Universidad de 
Granada, 18071 Granada (Spain), cabia@goliat.ugr.es\\
  $^2$ Institut d'Estudis Espacials de Catalunya/CSIC, Gran Capit\`a, 2-4,
  08034 Barcelona (Spain), isern@ieec.fcr.es\\
$^3$ IRAM, Avda. Divina Pastora 7, N\'ucleo Central, 18012 Granada (Spain)}
\date{Accepted . Received}
\pagerange{\pageref{firstpage}--\pageref{lastpage}}
\pubyear{}

\def\LaTeX{L\kern-.36em\raise.3ex\hbox{a}\kern-.15em
    T\kern-.1667em\lower.7ex\hbox{E}\kern-.125emX}

\begin{document}

\label{firstpage}

\maketitle

\begin{abstract}
The observed Be and B relationships with metallicity 
clearly support the idea that both elements have a primary origin and 
that are produced by the same class of objects. Spallation by  
particles accelerated during gravitational supernova events (SNII, 
SNIb/c) seems to be a likely origin. We show, in the context of a model of 
chemical evolution, that it is possible to solve the Li, Be and B 
abundance puzzle with the yields recently proposed by Ramaty et al. (1997) 
provided that SNII are unable to significantly accelerate helium nuclei and 
that different mechanisms are allowed to act simultaneously.  

\end{abstract}


\section{Introduction}
The origin of the light elements is still an open question in Astrophysics.
It is widely accepted that standard Big Bang nucleosynthesis can only produce
$^7$Li. The primordial abundance produced of this isotope is one order of 
magnitude below Solar System values and roughly coincides with that observed 
by Spite \& Spite (1982) in the hot halo dwarfs ({\it the lithium plateau}). 
For this reason, it is generally accepted that such an abundance is 
representative of the primordial nucleosynthesis and that, since the LiBeB 
isotopes do not have a cosmological origin, they must be created by 
galactic activity. 

It is well known that the interaction of cosmic rays (CR) with the
interstellar medium (ISM) can play an important role in the 
production of the LiBeB isotopes (Reeves, Fowler \& Hoyle 1970). In fact, the 
bulk of the $^6$Li, $^9$Be and $^{10}$B Solar System abundances can be 
naturally accounted with the standard model of galactic cosmic ray (GCR) 
nucleosynthesis. In this model, high energy protons and alpha particles 
collide with heavier nuclei (CNONe) present in the ISM to produce the light 
element isotopes (Meneguzzi, Audouze \& Reeves 1971). This model fails, 
however, to explain the present $^7$Li abundance and the Solar System 
$^7$Li/$^6$Li and $^{11}$B/$^{10}$B ratios. Furthermore, during the last 
decade, many observational studies have shown a linear relationship 
between Be and B abundances in metal-poor stars and the metallicity ([Fe/H]) 
(Rebolo et al. 1988; Gilmore et al. 1992; Boesgaard 1995; Molaro et al. 1997; 
Duncan et al. 1997). This {\it primary} behaviour cannot be easily explained 
by the standard GCR model: e.g Prantzos et al. (1993) used an ad-hoc 
hypothesis concerning the time evolution of the 
CR's escape-length, Abia et al. (1995) 
introduced artificial time dependences of the CR flux with the metallicity 
and the star formation rate and Casuso \& Beckman (1997) considered 
differential astration [see also Tayler (1995) and Yoshii, Kajino
\& Ryan (1997)]. The reason for using these hypotheses is that in the standard 
GCR model the LiBeB production rates are proportional to the global 
metallicity of the ISM and to the CR flux. Since the latter is assumed to be 
proportional to the supernova rate it is, in consequence, also proportional 
to the production rate of metals in the galaxy. Such a dependence would 
predict a slope of about two in the B and Be relationships with [Fe/H] rather 
than a slope of one as the observations show. Furthermore, Duncan et al. 
(1997) and Garc\'\i a-L\'opez et al. (1998) have recently shown a nearly 
constant B/Be ratio of $\sim 20$ in dwarf stars for a wide metallicity 
range, although the uncertainties in this ratio are important\footnote{This 
B/Be value is obtained taking into account N-LTE effects in the derivation 
of B and Be abundances}. 
This value is still compatible with the idea of a spallative origin of 
Be and B (e.g Fields, Olive \& Schramm 1994). Since this ratio is similar 
to that observed in the Solar System (Anders \& Grevesse 1989) it cannot 
have experienced large variations during the galactic evolution, 
which strongly supports the idea of a similar origin for Be and B.

A straightforward interpretation of these results is that the net production
rate of Be and B does not depend on the metallic abundance in the ISM, i.e.
the light-element production is not dominated by protons and alpha particles 
colliding with CNO nuclei but by these nuclei colliding with ambient protons 
and alpha-particles, probably in regions of massive star formation heavily 
enriched in these nuclei. The $\gamma$-ray observations from the Orion 
nebulae (Bloemen et al. 1994) provide additional support to this point of 
view since they are consistent with line emission from $^{12}$C$^*$ and 
$^{16}$O$^*$ produced by a large flux of low-energy ($<100$ MeV/nucleon) 
nuclei enriched in C and O. This might represent the first evidence of the 
existence of a considerable low-energy component in the spectrum of CRs 
(at least locally), as was suggested by different authors (Meneguzzi, 
Audouze  \& Reeves 1971; Canal, Isern \& Sanahuja 1980). Since the ejecta of 
gravitational supernovae (type II, Ib/c) naturally match the above conditions, 
these objects have been proposed as preferential sites for LiBeB spallation 
production (Gilmore et al. 1992). The ejecta in these explosions are indeed 
heavily enriched in CO nuclei and, under some conditions (type Ib/c 
supernovae), the concentration of these nuclei might even exceed that of H 
and/or He. Because the yield of CO nuclei in supernovae is almost independent 
of the metallicity of the progenitor star, the LiBeB produced by spallation 
during supernova outcomes would have a primary character as is observed
for Be and B.

In fact, within the framework of a galatic evolutionary model,
Vangioni-Flam et al. (1996) studied the production of Be and B, assuming
a low-energy spectrum  of the form $\rm{q(E)\sim E^{-n}}$, with $n=9$, 
and constant for $\rm{E\leq 30~MeV/n}$ in the CRs associated with Orion-like 
regions. They were able to explain the observed behaviour of Be and B vs. 
[Fe/H], and they obtained upper and lower limits for the contribution of 
this mechanism to the galactic evolution of Be and B abundances. Their results 
show that this mechanism might contribute up to $70\%$ of the observed Be 
and B abundances and that standard GCR nucleosynthesis is not the main 
source of $^9$Be and B in the galaxy. Recently, Ramaty et al. (1997; 
hereafter RKLR) studied the influence of different spectra and chemical 
compositions in the CRs produced by supernova on the LiBeB production. 
They showed from energetic 
arguments that due to the amount of Be necessary to account for its linear 
behaviour with [Fe/H], the CNO-rich, He-poor and H-poor CR source compositions 
are favoured. The observed Be/Fe ratio requires the investment of about 
$3\times 10^{49}$ to $2\times 10^{50}$ erg per gravitational supernova in 
these metallic CR, depending on whether or not H and He are accelerated with 
metals. Similar arguments led them to conclude that these CRs should have a 
hard-energy spectrum extending up to at least 50 MeV/n. 
From the constancy of the observed Be/Fe ratio and metallicity, they
also derived the necessary Be yield per supernova that is needed.
Assuming a $^{56}$Fe yield per SNII (Woosley \& Weaver 1995) of 
$\sim 0.11$ M$_\odot$, they obtained a Be yield of $2.8\times 
10^{-8}$ M$_\odot$ (within a factor of two of uncertainty from the observed 
Be/Fe ratio), almost irrespective of the progenitor star metallicity.

In this paper we have assumed that the Be yield from RKLR is
representative of the Be produced per gravitational supernova. Since this 
yield automatically sets the corresponding Li and B yields for a given 
CR spectrum and chemical composition, we use this to study the 
impact of such objects on the galactic evolution of the light element 
abundances. We discuss the results in the framework of different scenarios 
for the progenitors of type II and Ib/c supernovae and possible mechanisms 
for the CR acceleration in such objects.

\section{The model}

We use the same evolutionary model and approximations for the solar
neighbourhood as in Abia, Canal \& Isern (1991), which assumes
an exponential unenriched infall with an e-folding time of 4.5 Gyr,
and age of the galaxy of 13 Gyr. The adopted initial mass function, 
assumed to be constant in space and time, is that from Scalo (1986) in 
the mass 
range $0.5\leq$ M/M$_\odot\leq 100$ and the stellar life-times are from 
Talbot \& Arnett (1971). With these assumptions, the main characteristics 
of the solar neighbourhood are well reproduced (age-metallicity relation, 
current fraction of gas, the G-dwarf problem etc.).

Concerning the supernova scenarios, we consider that type Ia supernovae are 
the outcome of the merging of two white dwarfs in a binary system, their 
rates being calculated as in Bravo, Isern \& Canal (1993). 
For type Ib and type II 
supernovae (see below for type Ic) we assume that they are the outcome of 
the gravitational collapse of massive stars (M$\geq 12$ M$_\odot$). It is 
currently accepted that SNIb are caused by the explosion of the most massive 
stars, those that have lost their H--rich envelope (probably Wolf--Rayet 
stars) and even, in some cases, the He--rich one. The outcome will be then a 
{\it metallic supernova} in the sense of a CNO-rich and H(He)-poor ejecta, 
which, according to RKLR, are the energetically favoured supernovae to 
produce LiBeB by spallation. With these assumptions, our model has to fit the
observed evolution of the [CNO/Fe] ratios with metallicity (McWilliam 1997) 
and the present supernova rates in our galaxy. Assuming that our galaxy is of 
the Sb morphological type with a luminosity of $2\times 10^{10}$ L$_\odot$, 
Capellaro et al. (1997) estimate for our galaxy $4\pm 1$ SNIa, $2\pm 1$ 
SNIb+c and $12\pm 6$ SNII per millenium, i.e. SNIa/SNIb+c/SNII=1/0.5/3 within 
a $50\%$ of uncertainty. Finally, if we assume a fixed Be yield per 
gravitational supernova, we have to reproduce the observed LiBeB abundance 
evolution, which becomes a prediction of the model.

Throughout the calculations we have assumed a primordial contribution to the
lithium abundance of $^7$Li/H=$1.72\times 10^{-10}$ (Bonifacio \& Molaro 
1997) and the contribution of the standard GCR model. We have included the 
contribution of GCRs using the approximations of the leaky-box model
(Meneguzzi, Audouze \& Reeves 1971). The flux and the spectrum of CRs
have been obtained assuming a spectrum $\rm{q(E)\propto  E^{-2.7}}$
at the source, a scape-length of $10$ gcm$^{-2}$ and He/H$=0.08$, both 
constant in time. We assume that the abundances of the CNO nuclei in the 
accelerated particles scale with the global metallicity (Z/Z$_\odot$) keeping 
the abundance ratios observed today in the CR's flux. The temporal 
evolution of the abundance of CNO targets in the ISM has also been taken 
into account. In fact, this
is a prediction of our model. Figures 1 and 2 show the
evolution of the Be and B abundances assuming the sole contribution of
GCRs and normalizing their spectra at the source to obtain
the present-day Be abundance. As expected, the predicted Be and B evolution 
is steeper than the observed one (continuous line) and the typical spallation 
B/Be$\sim 10$ value and the $^{11}$B/$^{10}$B ratio which is always 
close to 2 are obtained (Figures 3 and 4, continuous line). Furthermore,
this model also fails to reproduce the present Li abundance (Figure 5, 
continuous line).

Now, we introduce the LiBeB production by spallation in gravitational
supernovae. We limit ourselves to the case in which the CR particles
accelerated in supernovae have a typical shock spectrum similar to that 
used by RKLR, which extends up to kinetic energies $\rm{E\sim 10~GeV/n}$. 
Once the spectrum has been fixed, and the yield of Be per supernova explosion 
adopted ($\sim 3\times 10^{-8}$ M$_\odot$), the Li and B yields only 
depend on the specific composition of the accelerated particles and ambient 
medium. If we assume that all gravitational supernovae (SNII and SNIb/c) 
contribute to the abundances of LiBeB isotopes, we have to consider 
two different chemical compositions (see Table 1):

Case 1: Ejecta representative of typical SNII (RKLR, model 2, Table 3).
This is obtained averaging, with the IMF as a weight, the chemical 
composition in the ejecta of supernovae from stars in the mass range 
12-40 M$_\odot$. The resulting composition resembles that of the present-day 
CR abundances except that they are less abundant in $^{12}$C and protons 
relative to $^{16}$O. The resulting Li and B yields per supernova normalized 
to that of Be are: $^6$Li/$^7$Li/$^9$Be/$^{10}$B/$^{11}$B=6.9/10/1/4/12.

Case 2: The chemical composition of the ejecta from metallic supernovae (RKLR, 
model 4, Table 3) is taken to be similar to that of the presently observed 
CR abundances except for H=He=0. In this case the relative yields are: 
$^6$Li/$^7$Li/$^9$Be/$^{10}$B/$^{11}$B=2.8/4.4/1/4/10.

\begin{table*}
\vbox to 220mm{\vfil
\caption{} \vfil}
\label{tab1len.tex}
\end{table*}

\section{Evolution of the light elements}

Figures 1 and 2 show the evolution of the Be and B abundances vs. [Fe/H]
respectively (short-dashed lines). In all cases it has been assumed that
$\rm{{SNIb+c/SNII}}\approx 0.2$, in agreement with observations.
A GCR contribution has also been included to reproduce the present Be
abundance. The evolution of the Be abundance is in perfect agreement with 
the observations while B seems slightly underproduced in the entire 
metallicity range.

\begin{figure}
\epsfxsize=9 cm
\epsfbox{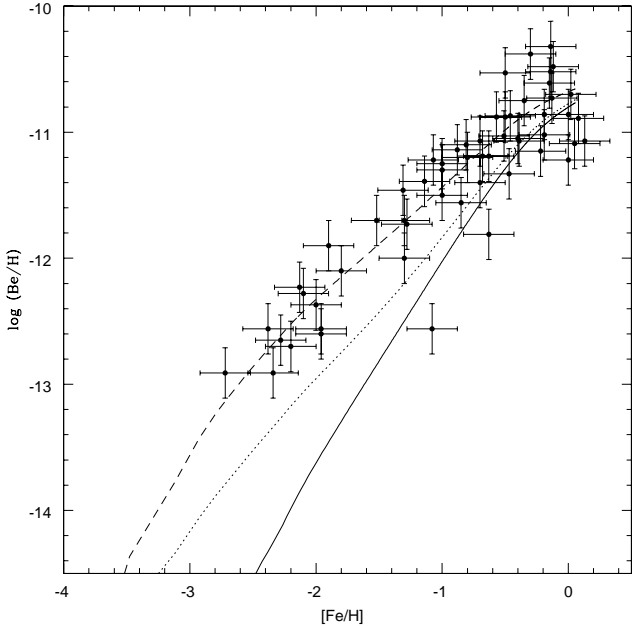}
\caption{Predicted Be/H vs. [Fe/H] evolution. Continuous line: GCR model 
alone. Short dashed line: GCR model plus case 1 + case 2. Dotted line: GCR 
model + case 2 and allowing a fraction $\rm A\sim 0.15$ of the massive stars 
in the range 12-40 M$_\odot$ to become SNIb/c (see text). Be abundances 
are taken from the literature (see text). Error bars are indicative.}
\end{figure}

\begin{figure}
\epsfxsize=9 cm
\epsfbox{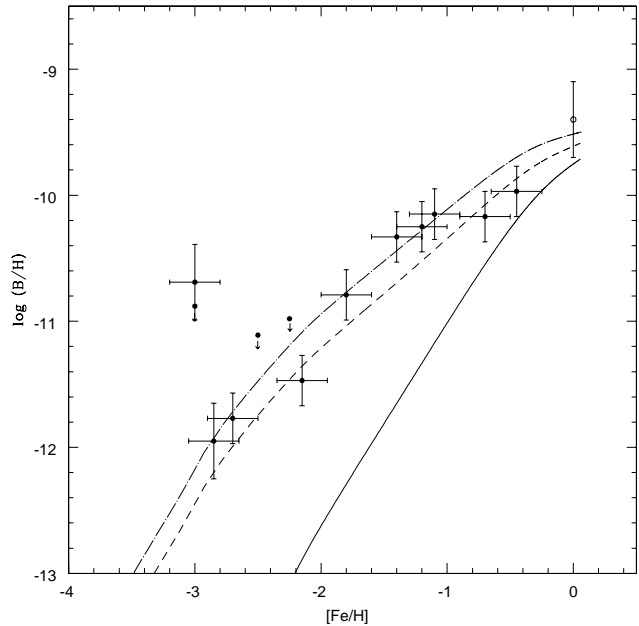}
\caption{Predicted evolution of B/H vs. [Fe/H]. Continuous line: GCR model
alone. Short dashed line: GCR model plus case 1 + case 2. Dash-dotted line: 
GCR + case 1 and case 2 + including a $^{11}$B production by neutrinos of 
$\sim 3.0\times 10^{-7}$ M$_\odot$ per gravitational supernova. Data are 
from Garc\'\i a-L\'opez et al. (1998).}
\end{figure}

Figures 3 and 4 show the evolution of the B/Be and $^{11}$B/$^{10}$B 
ratios (short--dashed lines). Because of the deficit in the B production, 
the evolution of the B/Be ratio is marginally consistent with observations 
at low metallicity. We obtain B/Be$\approx 14$ instead of 20 as 
suggested by the observations (Duncan et al. 1997; Garcia--L\'opez et al. 
1998). Furthermore, this value decreases to $\sim 10$, which is 
clearly smaller than the Solar System ratio\footnote{It is worth noting that 
the solar B abundance is still uncertain, varying from B/H$\approx 2\times 
10^{-10}$ to $7\times 10^{-10}$ in carbonaceous chondrites (Anders \& 
Grevesse 1989), to the photospheric value of $4\times 10^{-10}$ (Khol et al. 
1977)}. This drop is due to the progressive importance of the GCR production. 
Both the B/Be and the $^{11}$B/$^{10}$B ratios
display a maximum around [Fe/H]$\sim -2.5$. The maximum values reflect the
average B/Be and $^{11}$B/$^{10}$B production ratios in gravitational
supernovae (see Table 1). As the metallicity increases these ratios
evolve to their GCR values. A similar peak was also obtained by 
Vangioni-Flam et al. (1996). Note however, that there is no evidence
for this maximum in the B/Be observations.

The failure to reproduce the B/H vs. [Fe/H] relationship and the B/Be
and $^{11}$B/$^{10}$B ratios strongly suggests the existence of an
additional source of $^{11}$B like neutrino spallation in gravitational
supernovae (Domogatski \& Nadyozhin 1977; Woosley \& Weaver 1995).
This mechanism also produces noticeable quantities of $^{7}$Li.
Since the yields are strongly dependent on the neutrino temperature 
(a question requiring further study), we have adopted the values necessary 
to fit the B observations. Figures 2 to 4 (dash--dotted lines) show 
the results of such a contribution for a $^{11}$B neutrino yield of 
$\sim 3\times 10^{-7}$ M$_\odot$ per gravitational supernova (note that all 
stars with M$\geq 12$ M$_\odot$ can produce $^7$Li and $^{11}$B by neutrino
spallation). This value is not inconsistent with the nominal value quoted 
by Woosley \& Weaver (1995) of $6.5\times 10^{-7}$ M$_\odot$ and is also 
in agreement with the production range estimated by RLKR of (2 to $7\times 
10^{-7}$ M$_\odot$ per supernova). With the above neutrino yield, the 
predicted B/H vs. [Fe/H] evolution agrees very well with observations as 
does the B/Be ratio in the whole metallicity range. Concerning the 
$^{11}$B/$^{10}$B ratio, we obtain a value higher than 4 at early times, 
although  this decreases to $\sim 3$ at [Fe/H]$\sim 0.0$. Again, as 
the time increases, this ratio approaches the GCR value of $\sim$ 2.
Nevertheless, we believe that it should not be difficult to get exactly 
the Solar System ratio of $^{11}$B/$^{10}$B$=4.05\pm 0.05$ (Chaussidon 
\& Robert 1995) at [Fe/H]$\approx 0.0$ just using other evolutionary 
models constructed with parameters that are still compatible with 
the observational constraints. In our case, to obtain the Solar System value 
we need to increase the $^{11}$B neutrino yield to $\sim 6\times 10^{-7}$ 
M$_\odot$ (still within the theoretical limits), 
but in this case $^{11}$B is overproduced (see Fig. 4 dotted line). 

\begin{figure}
\epsfxsize=9 cm
\epsfbox{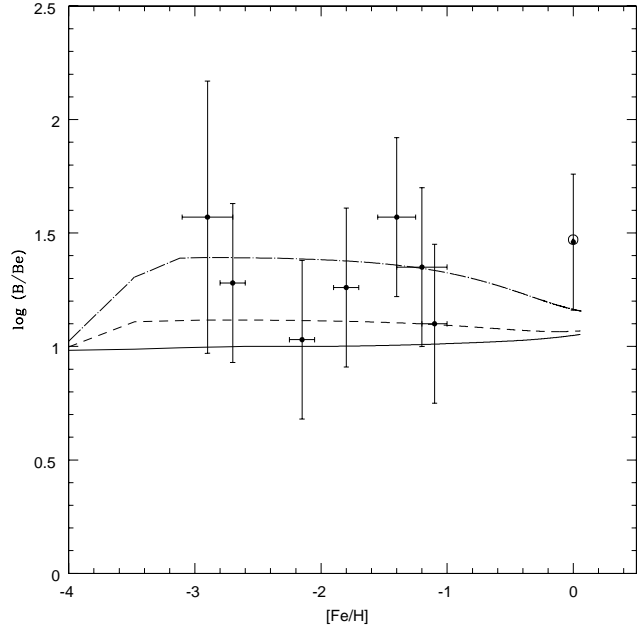}
\caption{Evolution of the B/Be ratio. Continuous line: GCR model olone. 
Short dashed line: GCR + case 1 and case 2. Dash-dotted line: GCR + case 1 
and case 2 with $\sim 3.0\times 10^{-7}$ M$_\odot$ of $^{11}$B production 
by neutrinos. Observed (B/Be)$_{\rm{N-LTE}}$ ratios are from Garc\'\i a
L\'opez et al. (1998). The data point at [Fe/H]=0.0 denotes the Solar
System ratio.}
\end{figure}

\begin{figure}
\epsfxsize=9 cm
\epsfbox{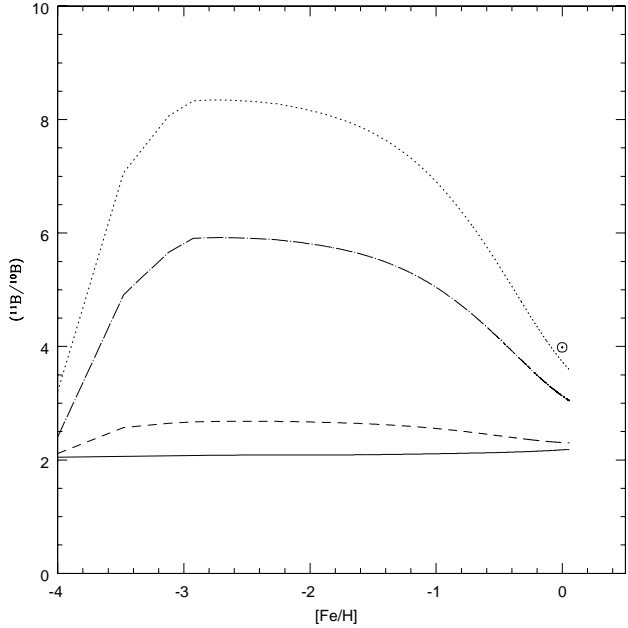}
\caption{Evolution of the $^{11}$B/$^{10}$B ratio. Continuous line: 
GCR model only. Short dashed line: GCR + case 1 and case 2 {\it without} 
$^{11}$B production by neutrinos. Dash-dotted line: GCR + case 1 and case 2 
{\it with} $^{11}$B production by neutrinos. Dotted-line: GCR + case 1 and 
case 2 but forcing a $^{11}$B production by neutrinos to match a ratio 
$\sim 4$ at [Fe/H]$\approx 0.0$ (see text). Dotted circle denotes the
Solar System ratio}
\end{figure}

Figure 5 displays the evolution of the Li abundance (dotted line) when 
an averaged $^7$Li yield due to neutrinos of $\sim 3\times 10^{-7}$ 
M$_\odot$ (Woosley \& Weaver 1995) per gravitational supernovae is included. 
The predicted evolution is compatible with the lithium plateau but fails 
to account for the present Li abundance and, in consequence, the Solar 
System ratio $^7$Li/$^6$Li$=12.5$\footnote{$^6$Li is not produced by
neutrino spallation}. Therefore an extra-source of $^7$Li is needed, 
probably with a longer lifetime than gravitational supernovae. The long 
dashed--dotted line of Figure 5 illustrates the behaviour 
of Li when the contribution of AGB stars in the mass range 1.5-8 M$_\odot$ is 
included according to the parametrization of Abia, Isern \& Canal (1995). In 
this case, we can adjust not only the evolution of the Li abundance, but we 
also obtain a $^7$Li/$^6$Li$=12.5$ ratio at the epoch of the Solar System 
formation. 

\begin{figure}
\epsfxsize=9 cm
\epsfbox{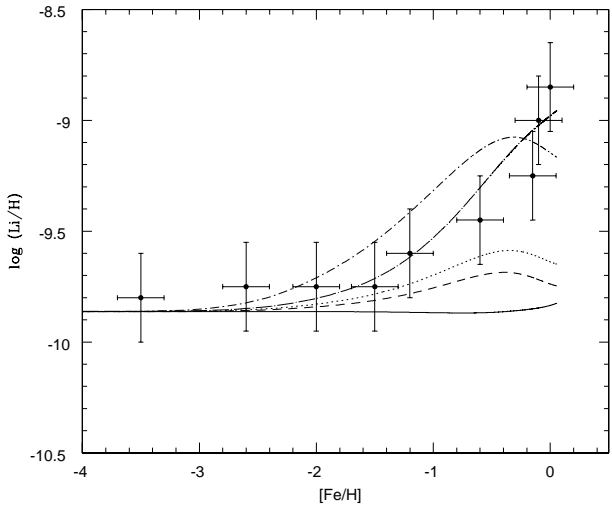}
\caption{Predicted evolution of the lithium abundance. Data represent the
upper envelope of the lithium abundances derived from hot dwarf stars 
(Rebolo, Molaro \& Beckmann 1988). Continuous line: GCR model only. 
Short dashed line: GCR + case 1 and case 2. Dotted line: the same as in the 
previous case but including neutrino spallation (see text). Long dash-dotted 
line: all the above cases plus the $^{7}$Li production in AGB stars 
(see text). Short dash-dotted line: GCR plus case 1 and case 2 but considering
CO-poor accelerated particles in SNII explosions (case 1).}
\end{figure}

The above situation presents, however, an important problem concerning the
acceleration of the particles in SNII. It seems quite difficult to accelerate 
the freshly nucleosynthesized matter (CNO nuclei) by the forward SNII shock, 
since the shock is formed ahead of the region containing the bulk of this 
matter. In this case, the most important spallation reactions for LiBeB 
production would be those between alpha-particles and protons colliding with 
alpha-particles and CNO nuclei forming part of the local circumstellar gas 
composition. Then, due to the low CNO abundances in CR and ambient 
medium, a {\it secondary} behaviour for Be and B is expected.
Feltzing \& Gustafsson (1994) showed that even assuming a high CNO abundance
in the circumstellar gas close to the supernova, very high local CR fluxes
($\geq 10^5$ cm$^{-2}$s$^{-1}$) are required in order to reproduce the 
Be and B linearity with metallicity. This would imply $\gamma$--ray fluxes
above those observed from supernova remnants and the Galaxy. One possible 
solution to this problem (as mentioned by RKLR) is to accelerate particles 
from matter freshly nucleosynthesized by the {\it reverse} SNII shock. 
Since this reverse shock will mainly affect the inner matter of the ejecta, 
the accelerated particles would be depleted of H and He. However, it seems 
improbable that this reverse shock could form and, even if it did, it would be
unlikely it could survive long enough to accelerate particles to
the appropriate energies for spallation reactions (Ellison, Drury 
\& Meyer 1997). Detailed studies are still needed to confirm these concepts. 

Since it seems that SNIb/c are the sole supernovae capable of accelerating the
CO nuclei with a suitable chemical composition, we have also examined the
minimum number of these supernovae needed to account for the observations. 
We assume that a fraction A of the stars that normally explode as SNII, 
i.e. those in the mass range 12-40 M$_\odot$, can also produce SNIb/c if 
they are members of close binary systems with the appropriate parameters 
(Nomoto, Iwamoto \& Suzuki 1995). However, to reproduce the 
Be/H vs. [Fe/H] relationship, A has to be $\sim 0.4$. Such a value of A 
means a very high efficiency of SNIb/c formation in close binary systems. In 
this case, the predicted ratio between SNIb+c and SNII rates would be a 
factor 3 higher than that observed in our galaxy, which seems quite improbable 
even taking into account the uncertainties in the determination of 
supernova rates. Nevertheless, the observed B evolution can still be 
reproduced using a reasonable value for $\rm A (\sim 0.15)$, but 
increasing the $^{11}$B neutrino yield per supernova to its 
estimated upper limit. However, in this case Be is clearly underproduced 
(see Fig. 1 dotted line). If this is the case, another primary source 
of $^{9}$Be must then be invoked. A similar result was obtained by 
Vangioni-Flam et al. (1996).

\section{Conclusions}

1) Type Ib/c supernovae seem to be key objects for the production of Be and B
by spallation. However, due to their low rates in the galaxy, an additional
contribution to the Be and B production by spallation due to type II
supernovae is needed. Using a reasonable shock spectrum for the CR source,
it is possible to explain the observed linear relation of Be and B 
abundances with metallicity although a neutrino contribution to B is necessary 
to account for the observed B/Be$\sim 20$. In this situation, the standard GCR 
model would play a minor role in the early evolution of the Be and B 
abundances. Additional sources of $^7$Li are still necessary to account 
for the evolution of its abundance. Objects with longer lifetime than 
supernova like AGB stars (and/or novae) seem to be necessary. A multi-source 
nature of this element is thus obvious.

2) This scenario has a serious problem with the acceleration of CO-rich 
and HHe-poor matter in SNII ejecta to sufficient energies for spallation 
reactions. It is not clear how a huge Li production
(due to $\alpha+\alpha$ reactions) in SNII relative to that of Be and 
B can be avoided if the acceletared particles are CO-poor. If that is
the case a production ratio Li/Be$\geq 100$ is predicted for a wide variety 
of CR spectra. This high Li production would be incompatible with the 
lithium plateau observed at low metallicity as shown in Figure 5 (short dash-
dotted line). 

3) The CR source spectrum shape associated with supernova explosions 
is another crucial point. It must extend to high kinetic energies 
(E$\geq 50$ MeV/n) in order for this scenario to be energetically plausible and
to avoid the overproduction of Li with respect to Be and B. Studies of LiBeB 
production using other CR spectra, not strict power laws, deserve special 
attention.

4) The existence of an additional source of $^{11}$B seems obligatory in order
to explain the Solar System's $^{11}$B/$^{10}$B$=4.05$ value. Production of 
$^{11}$B by neutrino spallation seems to be the best candidate. In this case, 
a $^{11}$B/$^{10}$B ratio higher than 4 at low metallicities is predicted, 
which might be tested by using very high resolution spectroscopy on 
very large telescopes (Rebull et al. 1997).

This work has been partially supported by CICYT grants 
PB96-1428 and ESP95-00091 and by a CIRIT grant.

\end{document}